# FAHP-based Mathematical Model for Exercise Rehabilitation Management of Diabetes Mellitus

Daoyan Pan [1,4], Kewei Wang [2], Zhiheng Zhou [3], Xingran Liu [3], Jie Shen [1,4]

（1.The Third Affiliated Hospital of Southern Medical University, Guangzhou 510630, Guangdong; 2.Guangdong Nanshan Pharmaceutical Innovation Research Institute, Guangzhou 510535,Guangdong; 3. School of Electronic and Information Engineering, South China University of Technology, Guangzhou 510640, Guangdong;4. Shunde Hospital of Southern Medical University, Shunde 528399,Guangdong）

**Abstract**: Exercise rehabilitation is an important part in the comprehensive management of patients with diabetes and there is a need to conduct comprehensively evaluation of several factors such as the physical fitness, cardiovascular risk and diabetic disease factors. However, special disease features of diabetes and its wide heterogeneity make it difficult to apply individualized approaches. In this study, a novel framework was established based on the Fuzzy Analytic Hierarchy Process (FAHP) approach to calculate various physiological factors weights when developing a diabetic exercise prescription. Proposed factors were investigated with respect to three groups which contains 12 different aspects. The relative weights were assessed by a database which established through a questionnaire survey. It is concluded that the physical fitness factors and cardiovascular risk factors need to be paid more attention to considered in the formulation of exercise rehabilitation programs than disease factors. And the cardiopulmonary function of physical fitness factors accounts for the highest importance. Furthermore, it was found that blood lipids have the lowest importance among studied factors. The mathematical model of exercise rehabilitation program for diabetes patients was established, which provided the theoretical basis for individualized guidance of exercise rehabilitation program.

**Keywords**: diabetes management；exercise rehabilitation；fuzzy analytic hierarchy process (FAHP)；mathematical model

# 1. Introduction

Diabetes has become an important public health problem around the world. Nearly 500 million people worldwide have diabetes. It is estimated that by 2035, the number of patients with diabetes will increase by 55%[1], The treatment cost of diabetes and its complications has become a heavy burden on society and families. Exercise rehabilitation is one of the preferred and important strategies in diabetes prevention and treatment programs. At present, various studies have confirmed that it can effectively reduce blood glucose, blood lipids, blood pressure and improve insulin sensitivity, which is conducive to the prevention and improvement of diabetes-related complications[2].

Previous diabetes exercise guidelines have made simple and general recommendations on intensity, frequency and duration of exercise rehabilitation for diabetic patients[3]. However, with the further in depth understanding of the particularity and heterogeneity of diabetes, it has been found that hyperglycemia, the increase of advanced glycation end products and the level of oxidative stress can cause damage to blood vessels, nerves, muscles of patients with diabetes and make them more vulnerable to suffer exercise-related injuries[4]. For example, diabetic patients are often complicated with cardiovascular autonomic neuropathy, diabetic retinopathy, diabetic peripheral neuropathy and other disease factors, which will greatly influence the formulation of exercise program and the safety of exercise rehabilitation therapy. Diabetic cardiovascular autonomic neuropathy leads to fast basic heart rate, exercise intolerance and low heart rate variability and severe cases can even induce asymptomatic myocardial infarction and sudden death[5]. Therefore, exercise rehabilitation training in the corresponding high-risk or diagnosed population needs to consider the impact of the above disease characteristics on exercise program formulation and exercise safety. At present, the simple, universal and one-size-fits-all Exercise rehabilitation program can no longer meet the individual needs of diabetic patients with different types of complications[6].

A scientific exercise rehabilitation program for diabetes needs to fully integrate the individual characteristics of diabetic patients, such as physical fitness factors, cardiovascular risk factors and diabetic disease factors, and effectively balance the safety and effectiveness at the same time. However, how to achieve this goals is still a difficult unsolved problem for the majority of diabetic exercise rehabilitation program makers[7].

In this study, we try to introduce fuzzy analytic hierarchy process (FAHP) to solve the above problems, FAHP can effectively integrate various physiological factors that need to be considered when formulating the exercise rehabilitation program for patients with diabetes, and through the establishment of fuzzy consistent matrix, to calculate various physiological factors weights. This method combines the idea of fuzzy mathematics with analytic hierarchy process, and relatively correct the subjective differences of human thinking in the process of making the program. FAHP has been applied in many fields, such as risk analysis, credit system construction and so on[8].

In view of the particularity of diabetes exercise rehabilitation management, there is little research on the comprehensive evaluation method of diabetes exercise rehabilitation program in the medical field, especially the application of fuzzy analytic hierarchy process in diabetes exercise rehabilitation management has not been reported. Therefore, this study attempts to apply fuzzy analytic hierarchy process to combine physical fitness factors, cardiovascular risk factors and diabetic disease factors and apply it to the formulation of exercise rehabilitation program for patients with diabetes. A comprehensive analysis of the weight proportion of each factor in the design of diabetes exercise rehabilitation program, so as to provide a solution for the individualization of the current clinical diabetes exercise rehabilitation program.

## 2. Construction of Evaluation Indexes and Evaluation Methods of Exercise Rehabilitation Program

### 2.1. Acquisition of Fuzzy Complementary Matrix

First of all, according to the demand of individualized exercise rehabilitation program for diabetes mellitus, combined with domestic and foreign guidelines and related researches[5–7], the analytic hierarchy process of diabetes exercise rehabilitation program was drawn (Table 1). The factors were classified into the following three groups: physical fitness factors (including age, BMI, cardiopulmonary function and cardiovascular disease); cardiovascular risk factors (including family history of sudden death, smoking, alcoholism and lack of exercise lifestyle); and diabetic disease factors (including blood glucose, blood pressure, blood lipid and diabetic complications). A questionnaire was used in order to determine the importance of above factors in affecting the diabetes exercise rehabilitation program.

Table 1. Evaluation index system of exercise rehabilitation program

| Target Layer | Criterion Layer | Index Layer |
|---|---|---|
| Formulate exercise rehabilitation program for diabetes(A) | Physical fitness factors(B1) | Age(C1) |
| | | BMI(C2) |
| | | Cardiopulmonary function(C3) |
| | | Cardiovascular disease(C4) |
| | Cardiovascular risk factors(B2) | Family history of sudden death(C5) |
| | | Smoking(C6) |
| | | Alcoholism(C7) |
| | | Lack of exercise lifestyle (C8) |
| | Diabetic disease factors(B3) | Blood glucose(C9) |
| | | Blood pressure(C10) |
| | | Blood lipid (C11) |
| | | Diabetic complications (C12) |

For the specific diabetic exercise program, the importance of the above factors varies, in order to accurately reflect the importance of factors in the design of diabetic exercise program, we use fuzzy analytic hierarchy process and combined with expert advice to determine the weight of these factors.

First of all, it is necessary to construct a fuzzy complementary judgment matrix. A questionnaire was designed and twenty-three experts were invited to fulfilled it. The importance order of the expert evaluation index in the traditional digital scale (Table 2) is transformed into the fuzzy complementary matrix according to the 0.1-0.9 scale method. The fuzzy consistent judgment matrix represents the comparison of the relative importance between the factors of the upper level and the related factors, assuming that the factors of the upper level are compared with the factors of the next level $a_1, a_2, \cdots, a_n$ is related, the A-B matrix B1, B2, B3 in the hierarchy diagram of Table 1. In order to avoid the interference of too many indexes to the expert judgment, we adopt a simplified method to establish the fuzzy complementary matrix. if the importance B1 > B2 > B3, the value is 0.9, 0.7, 0.5 in descending order of importance, as shown in Table 3.

Table 2. The quantitative scales and their fuzzy numbers

| Scale value | Definition | Description |
| --- | --- | --- |
| 0.5 | Equally important | Compared with the two factors, it is equally important |
| 0.6 | Moderately more important | Compared with the two factors, the former is moderately more important than the latter. |
| 0.7 | Strongly more important | Compared with the two factors, the former is strongly more important than the latter. |
| 0.8 | Very Strongly more important | Compared with the two factors, the former is very strongly more important than the latter. |
| 0.9 | Extremely more important | Compared with the two factors, the former is extremely more important than the latter. |
| 0.1，0.2，0.3，0.4 | Inverse comparison | If the judgment $r_{ij}$ is obtained by comparing factor $a_i$ with factor $a_j$, then the judgment obtained by comparing factor $a_j$ with factor $a_i$ is $r_{ji} = 1 - r_{ij}$ |

Table 3. Simplified method for establishing complementary matrix

|    | B1  | B2  | B3  |
|----|-----|-----|-----|
| B1 | 0.5 | 0.9 | 0.7 |
| B2 | 0.1 | 0.5 | 0.7 |
| B3 | 0.3 | 0.3 | 0.5 |

The value of the remaining B1-C, B2-C, B3-C matrix in descending order of importance is 0.8, 0.7, 0.6, and 0.5.

The fuzzy consistent judgment matrix can be expressed as follows (Table 4):

Table 4. Fuzzy consistent judgment matrix

| C     | $a_1$    | $a_2$    | …   | $a_n$    |
|-------|----------|----------|-----|----------|
| $a_1$ | $r_{11}$ | $r_{12}$ | …   | $r_{1n}$ |
| $a_2$ | $r_{21}$ | $r_{22}$ | …   | $r_{2n}$ |
| …     | …        | …        | …   | …        |
| $a_n$ | $r_{n1}$ | $r_{n2}$ | …   | $r_{nn}$ |

Among them, 0 《$r_{ij}$ 《1, $r_{ii}$=1, $r_{ij}$+$r_{ji}$=1, In order to give a more quantitative description of the comparison of importance between any two factors, the following settings are made: $r_{ij}$ indicates the importance of factor $a_i$ and factor $a_j$ relative to factor C, factor $a_i$ and factor $a_j$ have Fuzzy membership relation.

Fuzzy consistency matrix is defined as r = ($r_{ij}$) n × n, and the weights of factors $a_1$, $a_2$, ..., $a_n$ are $w_1$, $w_2$, …, $w_n$, respectively.

## 2.2. Consistency Test of Fuzzy complementary Matrices

The consistency index is often used to check the consistency of fuzzy matrix.

Let $w = (w_1, w_2, \cdots, w_n)^T$ be the weight vector of fuzzy matrix A, which satisfies the following formula:

$$w_i = \frac{\frac{n}{2} - 1 + \sum_{j=1}^{n} a_{ij}}{n(n-1)}$$

Among them

$$\sum_{i=1}^{n} w_i = 1, w_i \geq 0, (i = 1, 2, \cdots, n)$$

Fuzzy complementary matrix A * is the weight matrix of fuzzy complementary matrix A, which satisfies the calculation formula as follows:

$$A_{ij}^* = \frac{w_i}{w_i + w_j}$$

Then the compatibility index of matrix A * and A $CI(A, A^*)$, is defined as:

$$CI(A, A^*) = \frac{\sum_i^n \sum_j^n |a_{ij} - a_{ij}^*|}{n^2}$$

Which is possible to be defined as consistency index of fuzzy complementary matrix A.

If $CI(A, A^*) \leq \rho$, the fuzzy complementary matrix A is defined to have satisfactory consistency, that is, it has passed the consistency test. The ρ value is usually set to 0.05 or 0.1, depending on the evaluator's subjective attitude. In this paper, 0.1 is used. The results are shown in Table 5.

Table 5. Consistency check

|  | CI | Consistency check |
| --- | --- | --- |
| B1-C | 0.0578 | Pass |
| B2-C | 0.0408 | Pass |
| B3-C | 0.0509 | Pass |
| A-B | 0.0216 | Pass |

### 2.3 Transform fuzzy complementary matrix into fuzzy consistency matrix

The fuzzy complementary matrix $A = (a_{ij})_{n \times n}$ is summed by row, shows as:

$$r_i = \sum_{j=1}^{n} a_{ij}, (i = 1, 2, \cdots, n)$$

Through the following mathematical transformation formulas:

$$r_{ij} = \frac{r_i - r_j}{2n} + \frac{1}{2}$$

Fuzzy matrix A can be transformed into fuzzy consistency matrix $R = [r_{ij}]_{n \times n}$.

Fuzzy consistency matrix R naturally satisfies the consistency condition, and there is no need to check its consistency. Because by definition, let the fuzzy complementary matrix R= [$r_{ij}$ ]_(n×n), for $\forall i, j, k$ ,if it satisfies following condition:

$$r_{ij} = r_{ik} - r_{jk} + 0.5, \quad (i, j, k = 1, 2, \cdots, n)$$

Then the matrix R is called fuzzy consistency matrix. The above matrix obviously satisfies the conditions.

### 2.4 Transform fuzzy consistency matrix into positive reciprocal matrix

Adopt transformation formula as follows:

$$e_{ij} = \frac{r_{ij}}{r_{ji}}$$

The fuzzy consistency matrix $R = [r_{ij}]_{n \times n}$ is transformed into the reciprocal matrix $E = [e_{ij}]_{n \times n}$, and the index weight is calculated.

### 2.5 Calculate the weight of factors

In this paper, the square root method is used to calculate the weight coefficient of the factors.

According to the square root formula as follows:

$$W_i = \frac{\left(\prod_{i=1}^{n} a_{ij}\right)^{\frac{1}{n}}}{\sum_{i=1}^{n} \left(\prod_{j=1}^{n} a_{ij}\right)^{\frac{1}{n}}}, \quad (i = 1, 2, \cdots, n)$$

The calculation is that the elements of the judgment matrix A are multiplied by rows to get a new vector, and then each component of the new vector is given to the n power, and then the weight vector is obtained by normalization.

According to the formula, the initial weight vector $w^{(0)}$ is defined as follows:

$$w^{(0)} = (w_1, w_2, \cdots, w_n)^T$$

$$= \left[\frac{\sqrt[n]{\prod_{i=1}^{n} r_{1j}}}{\sum_{i=1}^{n} \sqrt[n]{\prod_{j=1}^{n} r_{ij}}}, \frac{\sqrt[n]{\prod_{i=1}^{n} r_{2j}}}{\sum_{i=1}^{n} \sqrt[n]{\prod_{j=1}^{n} r_{ij}}}, \cdots, \frac{\sqrt[n]{\prod_{i=1}^{n} r_{nj}}}{\sum_{i=1}^{n} \sqrt[n]{\prod_{j=1}^{n} r_{ij}}}\right]^T$$

Taking the weight vector $w^{(0)}$ as the iterative initial value of the eigenvalue method, that is, $V_0 = (v_{0.1} v_{0.2} \cdots, v_{0.n})^T = w^{(0)}$ is the iterative initial value, the

following iterative formula is used:
$$V_{K+1} = EV_K$$

Calculate $V_{K+1}$ and $\|V_{k+1}\|_\infty$, when $|\|V_{k+1}\|_\infty - \|V_k\|_\infty| < \varepsilon$, where $\varepsilon$ is the iterative accuracy, this paper takes 0.0001, then $\|V_{k+1}\|_\infty$ is $\lambda\_max$, and $V_{K+1}$ is normalized.

$$V_{k+1} = \left[\frac{v_{k+1,1}}{\sum_{i=1}^n v_{k+1,i}}, \frac{v_{k+1,2}}{\sum_{i=1}^n v_{k+1,i}}, \cdots, \frac{v_{k+1,n}}{\sum_{i=1}^n v_{k+1,i}}\right]^T$$

$w^{(k)} = V_{k+1}$ is the final vector, and the iteration ends, Otherwise, according to the following iterative formula:

$$V_{k+1} = \left[\frac{v_{k+1,1}}{\|V_{k+1}\|_\infty}, \frac{v_{k+1,2}}{\|V_{k+1}\|_\infty}, \cdots, \frac{v_{k+1,n}}{\|V_{k+1}\|_\infty}\right]^T$$

Taking $V_{k+1}$ as a new initial iteration value, it enters the next iteration cycle and iterate so that the maximum weight vector $w^{(k)}$ is obtained, where k is the number of iterations. Finally, the comprehensive weights of each level are calculated, and the comprehensive weights of each index are obtained.

## 3. Results

In this study, the questionnaires of 23 clinical experts were collected, and all the questionnaires were completed. After establishing the fuzzy matrix and iterating, the weights of each level were finally obtained (Table 6).

Table 6 weight of each level

|      | w1       | w2       | w3       | w4       |
|------|----------|----------|----------|----------|
| B1-C | 0.229247 | 0.19914  | 0.30088  | 0.270733 |
| B2-C | 0.294592 | 0.236773 | 0.216952 | 0.251683 |
| B3-C | 0.273183 | 0.259186 | 0.187154 | 0.280477 |
| A-B  | 0.356635 | 0.338477 | 0.304887 |          |

A-B is multiplied by B1-C, B2-C and B3-C to get the comprehensive weight of each index (Table 7)

Table 7 Global weights for each factor

| Factors | Weight | Subfactors | Global weight |
|---|---|---|---|
| Physical fitness factors | 35.7% | Age | 8.18% |
| | | BMI | 7.10% |
| | | Cardiopulmonary function | 10.73% |
| | | Cardiovascular diseases | 9.66% |
| Cardiovascular risk factors | 33.8% | Family history of sudden death | 9.97% |
| | | Smoking | 8.01% |
| | | Alcoholism | 7.34% |
| | | Lack of exercise lifestyle | 8.52% |
| Diabetic disease factors | 30.5% | Blood glucose | 8.33% |
| | | Blood pressure | 7.90% |
| | | Blood lipid | 5.71% |
| | | Diabetic complications | 8.55% |

## 4 Discussion

On the basis of combining the guidelines for exercise rehabilitation of diabetes, this study first applied the FAHP method to the formulation of exercise rehabilitation program for diabetes, and systematically analyzed the weight of physical fitness factors, cardiovascular risk factors and diabetic disease factors in the formulation of exercise rehabilitation program for diabetes by means of expert score. The purpose of this study is to provide a new solution for the individualized and accurate formulation of the follow-up exercise rehabilitation program for diabetes.

Our study found that physical fitness factors and cardiovascular risk factors need to be paid more attention to and considered in the formulation of exercise rehabilitation programs than diabetic disease factors. Our results suggest that the cardiopulmonary function of physical fitness factors accounts for the highest weight, suggesting that more consideration should be given to the cardiopulmonary function of individuals with diabetes when formulating exercise rehabilitation programs for diabetes. Cardiopulmonary function is the fundamental condition and basis for the implementation of exercise rehabilitation programs. Guidelines also suggest that cardiopulmonary function assessment should be carried out first in some patients with diabetes, especially those with a long course of diabetes and a history of cardiovascular disease, in order to develop an appropriate exercise program to ensure the safety and effectiveness of exercise therapy, which is consistent with the results of this study.

In addition, our study also suggests that the course of diabetes is also an important consideration in formulating exercise rehabilitation programs for patients with diabetes, and the main reason may be the particularity and heterogeneity of diabetes. For example, diabetic complications will affect the efficacy and safety of exercise rehabilitation. For patients with diabetic retinopathy, high-intensity exercise (75% VO2max), high-impact jumping and Valsalva movements should be avoided. Type 2 diabetes (especially patients with peripheral neuropathy and peripheral vascular disease) should have regular foot examinations before and after exercise intervention, and patients with active foot injuries or ulcers should avoid weight-bearing exercise[9].

At present, with the in-depth study of the pathogenesis of diabetes and the accumulation of many large-scale clinical evidence-based medicine data, the concept of diabetes management focusing on cardiovascular risk factors on the basis of comprehensive management of multiple factors has been gradually accepted by the public. In fact, patients with diabetes are 2-4 times more likely to suffer from cardiovascular disease than healthy people[10]. Sedentary and other inactive lifestyles have been shown to significantly increase the risk of a range of chronic diseases. In addition, poor lifestyles such as lack of activity, smoking and alcohol abuse are one of the greatest risk factors for chronic diseases and early death. They are involved in the occurrence and development of sudden death, acute myocardial infarction, stroke, sarcopenia and various complications. The cardiovascular health level of most stroke survivors is low, and the partial limitation of mobility can lead to serious consequences such as stroke recurrence. Our research suggests that cardiovascular disease history, family history of sudden death and other risk factors such as smoking, alcoholism and lack of exercise all occupy a high weight. This series of indicators highlight the importance of safety in the formulation of diabetes exercise rehabilitation programs. Therefore, the formulation of sports rehabilitation programs for this part of the population needs to pay more attention to the above contents, individualized exercise rehabilitation programs[11]. For example, for patients who lack exercise, avoid directly engaging in unaccustomed strenuous physical exertion and high competitive exercise, advocate proper warm-up and post-exercise stretching exercise, and emphasize strict compliance with the prescribed target heart rate and a more cautious "progressive transitional exercise program."[12]

With the increase of age, the functions of most physiological systems of the human body will gradually decline. These age-related physiological changes widely affect the function of tissues and organs of patients, and then have an impact on the exercise ability of patients. For example, physiological indexes such as maximum oxygen uptake (VO2max) and muscle exercise ability will decrease with physiological aging, and these indicators are important indicators for the formulation of exercise program. Therefore, it is necessary to fully consider the age factor when formulating the exercise prescription and strike a balance between efficacy and safety[13].

Scientific exercise program can effectively control blood glucose, which plays an important and positive role in promoting the compliance of diabetic patients with exercise therapy. However, the current research is not consistent with the effects of different forms, time and intensity of exercise on blood glucose, while the adjustment of diet and antidiabetic drugs can have an immediate effect. Our research shows that the proportion of blood glucose factor in the formulation of individualized exercise rehabilitation program is low, which is consistent with the relevant guidelines, that is, exercise rehabilitation should not be limited if blood glucose can be controlled under the taboo exercise threshold[14].

In the comprehensive control goal of patients with diabetes, the level of blood lipid is an extremely important factor to evaluate the cardiovascular risk factors of diabetes. At present, numerous studies have found that exercise training can make favorable changes of triglyceride, total cholesterol and low density lipoprotein cholesterol in the blood[15,16]. Other studies have shown that resistance training can effectively increase the level of high density lipoprotein cholesterol in the blood[17]. In this study, it is found that the blood lipid level accounts for a relatively low proportion in the formulation of diabetic exercise rehabilitation program. The possible reason may related to the long time period for the improvement of blood lipid through exercise therapy. At present, the comprehensive management of diabetic patients are emphasized, including the control of blood lipids, and related guidelines emphasize the importance of using statins in diabetic patients with dyslipidemia. Oral lipid-lowering drugs can make blood lipids reach the standard faster and better, so the level of blood lipids in patients with diabetes may not be a priority compared with other indicators in the formulation of diabetes exercise rehabilitation program.

In view of the low rate of reaching the standard of blood glucose and the high incidence of complications in patients with diabetes, it is very important to formulate a reasonable, effective and safe exercise program for the huge population of diabetic patients around the world. We have made a preliminary exploration on the exercise rehabilitation program for diabetes based on FAHP, and provided ideas for the scientific and personalized customization of the relevant exercise programs, but this study still has a little deficiency in the comprehensiveness of the index system, the coverage of data collection, including the professional attributes of experts, and the discrimination of some weight is not good. The coefficient will be corrected through the accumulation of actual clinical data.